\documentclass[letter]{aa}  
\usepackage{graphicx,epstopdf}
\usepackage{txfonts}
\usepackage{float}
\usepackage{flafter}
\usepackage{amsmath}
\usepackage{amssymb}
\usepackage{booktabs}
\usepackage{multirow}
\usepackage{hhline}
\usepackage{enumitem}
\usepackage[normalem]{ulem}
\usepackage{longtable}
\usepackage[section]{placeins}
\usepackage{placeins}
\usepackage{hyperref}

\usepackage{savesym}
\savesymbol{tablenum}
\usepackage{siunitx}
\restoresymbol{SIX}{tablenum}

\DeclareSIUnit\parsec{pc}
\DeclareSIUnit\mpc{Mpc}
\DeclareSIUnit\year{yr}
\DeclareSIUnit\Gyr{Gyr}
\DeclareSIUnit\Jy{Jy}
\DeclareSIUnit\mJy{mJy}
\DeclareSIUnit\beam{beam}
\DeclareSIUnit\mK{mK}

\newcommand{\gdr}{$\delta_{ \rm GDR}$}

\newcommand{\s}{{\it Spitzer}}
\newcommand{\h}{{\it Herschel}}
\newcommand{\msol}{$\rm M_{\odot}$}
\newcommand{\md}{$M_{\rm d}$}

\newcommand{\lir}{$L_{\rm IR}$}
\newcommand{\ms}{$M_{\ast}$}

\newcommand{\td}{$T_{\rm d}$}
\newcommand{\mgas}{$M_{\rm gas}$}
\newcommand{\Mstar}{$M_{\ast}$}

\newcommand{\ci}{[C\,{\footnotesize I}]}

\newcommand{\cione}{[C\,{\footnotesize I}]$(^3P_1\,-\,^{3}P_0)$}
\newcommand{\citwo}{[C\,{\footnotesize I}]$(^3P_2\,-\, ^{3}P_1)$}

\newcommand{\lci}{$L^{\prime}_{[\rm CI]({^3}{\rm P}{_1}-{^3}{\rm P}{_0})}$}
\newcommand{\lcitwo}{$L^{\prime}_{[\rm CI]({^3}{\rm P}{_2}-{^3}{\rm P}{_1})}$}

\graphicspath{{./}{pics/}}

\begin{document} 

\title{Deceptively cold dust in the massive starburst galaxy GN20 at $z\sim4$}

\author{Isabella Cortzen\thanks{E-mail: cortzen@nbi.ku.dk}\inst{1,2}\and Georgios E. Magdis\inst{1,2,3}\and Francesco Valentino\inst{1,2}\and Emanuele Daddi\inst{4} \and Daizhong Liu\inst{5} \and Dimitra Rigopoulou\inst{6} \and Mark Sargent\inst{7} \and Dominik Riechers\inst{8} \and Diane Cormier\inst{4}\and Jacqueline A. Hodge\inst{9}\and Fabian Walter\inst{5} \and David Elbaz\inst{4}\and Matthieu B\'ethermin\inst{11}\and Thomas R. Greve\inst{12,1}\and Vasily Kokorev\inst{1,2}\and Sune Toft\inst{1,2}}

\institute{Cosmic Dawn Center (DAWN) \and
Niels Bohr Institute, University of Copenhagen, Blegdamsvej 17, DK-2100 Copenhagen \and
DTU-Space, Technical University of Denmark, Elektrovej 327, DK-2800 Kgs. Lyngby \and 
CEA, IRFU, DAp, AIM, Universit\'e Paris-Saclay, Universit\'e Paris 
Diderot, Sorbonne Paris Cit\'e, CNRS, F-91191 Gif-sur-Yvette, France \and
Max Planck Institute for Astronomy, Königstuhl 17, D-69117 
Heidelberg, Germany \and
Department of Physics, University of Oxford, Keble Road, 
Oxford OX1 3RH, UK \and 
Astronomy Centre, Department of Physics and Astronomy, 
University of Sussex, Brighton, BN1 9QH, UK \and
Department of Astronomy, Cornell University, Space Sciences 
Building, Ithaca, NY 14853, USA \and 
Leiden Observatory, Leiden University, P.O. Box 9513, 2300 
RA Leiden, The Netherlands \and
Max–Planck Institut für Astronomie, Königstuhl 17, D-69117 
Heidelberg, Germany \and 
Aix Marseille Univ., Centre National de la Recherche Scientifique, 
Laboratoire d'Astrophysique de Marseille, Marseille, France \and
Department of Physics and Astronomy, University College London, 
Gower Street, London WC1E 6BT, UK}

\authorrunning{I. Cortzen et al.}

\date{\today}
\date{Accepted XXX. Received YYY; in original form ZZZ}


\label{firstpage}

\abstract{We present new observations, carried out with IRAM NOEMA,  of the atomic neutral carbon transitions 
[\ion{C}{I}]$(^3P_1$--$^3P_0)$ at 492\,GHz and [\ion{C}{I}]$(^3P_2$--$^3P_1)$ at 809\,GHz 
of GN20, a well-studied star-bursting galaxy at $z=4.05$. The high luminosity line ratio [\ion{C}{I}]$(^3P_2$--$^3P_1)$
/[\ion{C}{I}]$(^3P_1$--$^3P_0)$  implies an excitation temperature of 
$48^{+14}_{-9}\,$\si{\kelvin}, which is significantly higher than the apparent dust temperature of 
\td=$33\pm2\,$\si{\kelvin} ($\beta=1.9$) derived under the common assumption of an optically 
thin far-infrared dust emission, but fully consistent with \td$=52\pm5\,$\si{\kelvin} of a 
general opacity model where the optical depth ($\tau$) reaches unity at a wavelength of 
$\lambda_0=170\pm23\,$\si{\micro\m}. Moreover, the general opacity solution returns a factor of 
$\sim 2\times$ lower dust mass and, hence, a lower molecular gas mass for a fixed gas-to-dust ratio, than with the optically thin dust model. The derived properties of GN20 thus provide an appealing 
solution to the puzzling discovery of starbursts appearing colder than main-sequence galaxies 
above $z>2.5$, in addition to a lower dust-to-stellar mass ratio that approaches the physical 
value predicted for starburst galaxies.}


\maketitle


\section{Introduction}\label{sec:intro}

Over the last decade, it has been established that the majority of star-forming 
galaxies (SFGs) fall into a tight correlation between the star formation rate (SFR) 
and the stellar mass (\Mstar), forming a "main-sequence" (MS) with a normalization 
that increases with redshift 
\citep[e.g.,][]{Brinchmann2004,Daddi2007,Noeske2007,Elbaz2007, Magdis2010}. 
Outliers of this relation are defined as starburst galaxies (SBs), existing at 
all redshifts. While the star formation in MS galaxies is governed by secular processes, 
merger-induced events or galaxy interactions are thought to trigger it in SBs 
\citep[e.g.,][]{Cibinel2019}.

In the interstellar medium (ISM), the thermal emission from dust grains heated by 
UV photons originating from newly formed stars dominates the spectral energy 
distribution (SED) of galaxies \citep[at $\sim 8-1000$\,\si{\micro\m},][]{Sanders1996}. 
Modeling of the rest-frame far-infrared (FIR) and the Rayleigh-Jeans (RJ) tail of the SED 
can be used to derive properties including the dust mass (\md), the infrared luminosity 
(\lir), the intensity of the radiation field \citep[$\langle$U$\rangle 
\propto$ \lir / \md :][]{Draine2007}, and the mass-weighted dust temperature 
(\td) where $\langle$U$\rangle=(T_{\rm d}/18.9)^{6.04}$ \citep{Magdis2012b,Magdis2017}.

With the ever-increasing number of galaxy populations with well-studied infrared properties, 
several puzzling findings have started to emerge, especially for high-redshift SBs. 
First, their dust-to-stellar mass ratios (\md$/$\ms) are found to be extremely large 
\citep[reaching 0.1:][]{Tan2014}, with a stellar mass budget that is unable to account 
for the inferred dust production \citep{Bethermin2015}. 
Second, while the intensity of the radiation field in MS galaxies rises with 
increasing redshift up to $z\sim4$ \citep{Magdis2017, Jin2019}, mirroring the increase 
in the specific star formation rate (sSFR=SFR/\ms) in the same time interval 
\citep[][for \td: \citet{Schreiber2018}]{Bethermin2015}, the evolution is less clear 
for SBs. While \citet{Schreiber2018} report a trend of increasing 
$T_{\rm d}$ with both redshift and offset from the MS, the latter, independently of redshift, 
\citet{Bethermin2015} observe no evolution of the mean radiation field (hence, dust temperature) 
with redshift for strong SBs with ${\rm sSFR}>10\times{\rm sSFR}_{\rm MS}$, which 
become apparently colder than MS galaxies at $z>2.5$, 
which is at odds with the expectations. A possible solution to the latter could be offered by 
a more general treatment of the modeling of the FIR emission that in the vast majority 
of the literature. Also, due to the limited sampling of the SEDs in the FIR to RJ regime, 
such modeling is performed under the assumption of optically thin FIR emission for both MS and SB 
galaxies. Indeed, observational studies of local ultra-luminous infrared galaxies 
(ULIRGs) and high-redshift massive SBs indicate that the dust could remain optically 
thick out to rest-frame $\lambda_{0}=100-200\,$\si{\micro\m} 
\citep[e.g.,][]{Blain2003a, Huang2014, Lutz2016, Spilker2016, Riechers2013, Hodge2016, Simpson2017}
and, in the most extreme case, out to millimeter wavelengths as reported for the star-bursting 
nucleus of Arp 220 \citep{Scoville2017Arp220}. 
If the FIR dust emission is optically thick, the suppressed continuum emission in the 
Wien's part of the IR emission shifts the peak of the SED to longer wavelengths, 
mimicking apparently cold \td, while, in fact, the actual luminosity-weighted \td\ of 
the sources would be considerably warmer. The main difficulty is that the optically 
thin or thick solutions are heavily degenerate; the same SED could arise from either cold 
and optically thin or a warm and optically thick FIR dust emission with no robust way to 
discriminate between the two by simply using continuum observations. An independent 
proxy for \td\ is, thus, required to break this degeneracy.

In this work, we present new Northern Extended Millimeter Array (NOEMA) observations of GN20, a 
well-known massive (stellar mass of \Mstar$\sim 10^{11}$\,\msol: \citealt{Tan2014}) starburst galaxy at 
$z=4.0553$ \citep{Pope2006, Daddi2009}, targeting both atomic neutral carbon lines, \cione\ and \citwo.
The simple three-level structure of the atom allows us to use the \ci\ line luminosity 
ratio to derive the excitation temperature ($T_{\rm ex}$), which was recently reported to correlate 
with \td\ derived assuming optically thin FIR dust emission on sub-galactic 
scales for 
nearby (U)LIRGs \citep{Jiao2019a, Jiao2019b}, suggesting that the gas probed by \ci\ and 
the dust are correlated on kpc scales. The \ci\ line ratio might thus be used as an 
independent empirical indicator of the dust temperature, potentially breaking the degeneracy 
between an optically thick and thin case for the FIR dust emission.

Throughout the paper, we adopt $H_{0} = 70$ \si{\km\per\s\per\mpc}, $\Omega_{\rm M} = 0.30$, 
$\Omega_{\Lambda} = 0.70$, and a \citet{Chabrier2003} initial mass function (IMF).\vspace{-2mm}

\section{Observations and data reduction} \label{sec:observations}
We used IRAM NOEMA to observe the \cione, \citwo, and CO(7-6) line transitions in the 
GN20 protocluster \citep{Daddi2009}. The observations took place in March 2017 using the 
D configuration for a total on-source time of 7.6 hours (program W16DZ, PI: G. Magdis).
The \cione\ line (rest frequency: $\nu_{\rm rest}=492.161$\,GHz) is redshifted to 
$\nu=97.355$\,GHz at $z=4.0553$ with a primary beam of 51.8\si{\arcsecond}. 
We set our pointing center to the coordinates of GN20 (RA: 12h37m11.89s, 
DEC: +62d22m12.1s) to detect the \ci\ and CO lines in this galaxy. Although the D-configuration 
leads to a relatively low spatial resolution ($\sim$3--6$''$), it is the most suitable 
configuration for a detection experiment as ours. As the observations of the two other GN20 
protocluster members, GN20.2a and GN20.2b, are affected by a primary beam attenuation of 
about $0.2-0.7$, no lines were detected and we could not derive any constraining measurements 
for these galaxies.

\begin{figure*}[!htb]
\begin{center}
    \includegraphics[width=.49\linewidth, trim={30 30 30 30}, clip]{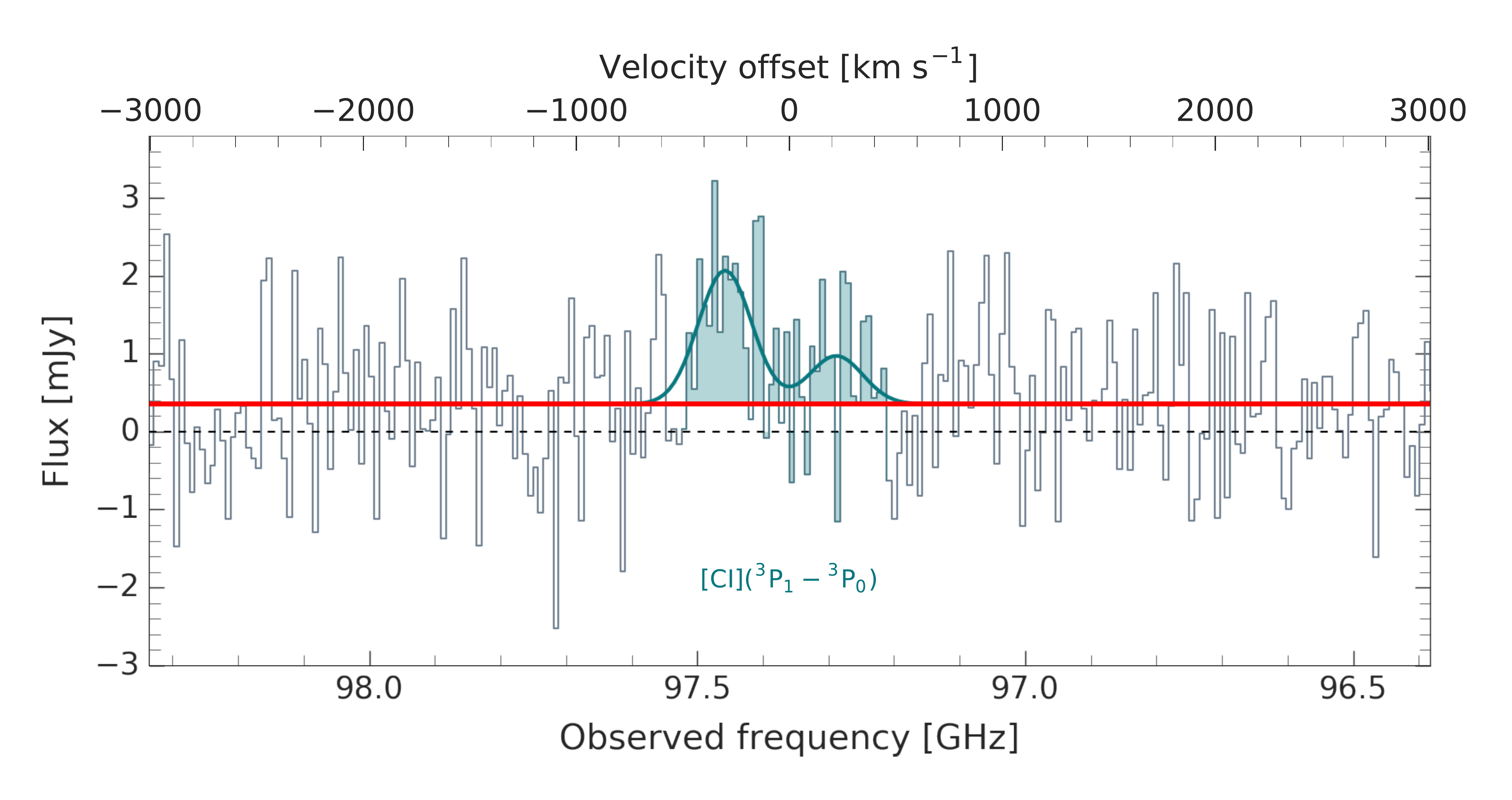}\quad\includegraphics[width=.49\linewidth, 
    trim={30 30 30 30}, clip]{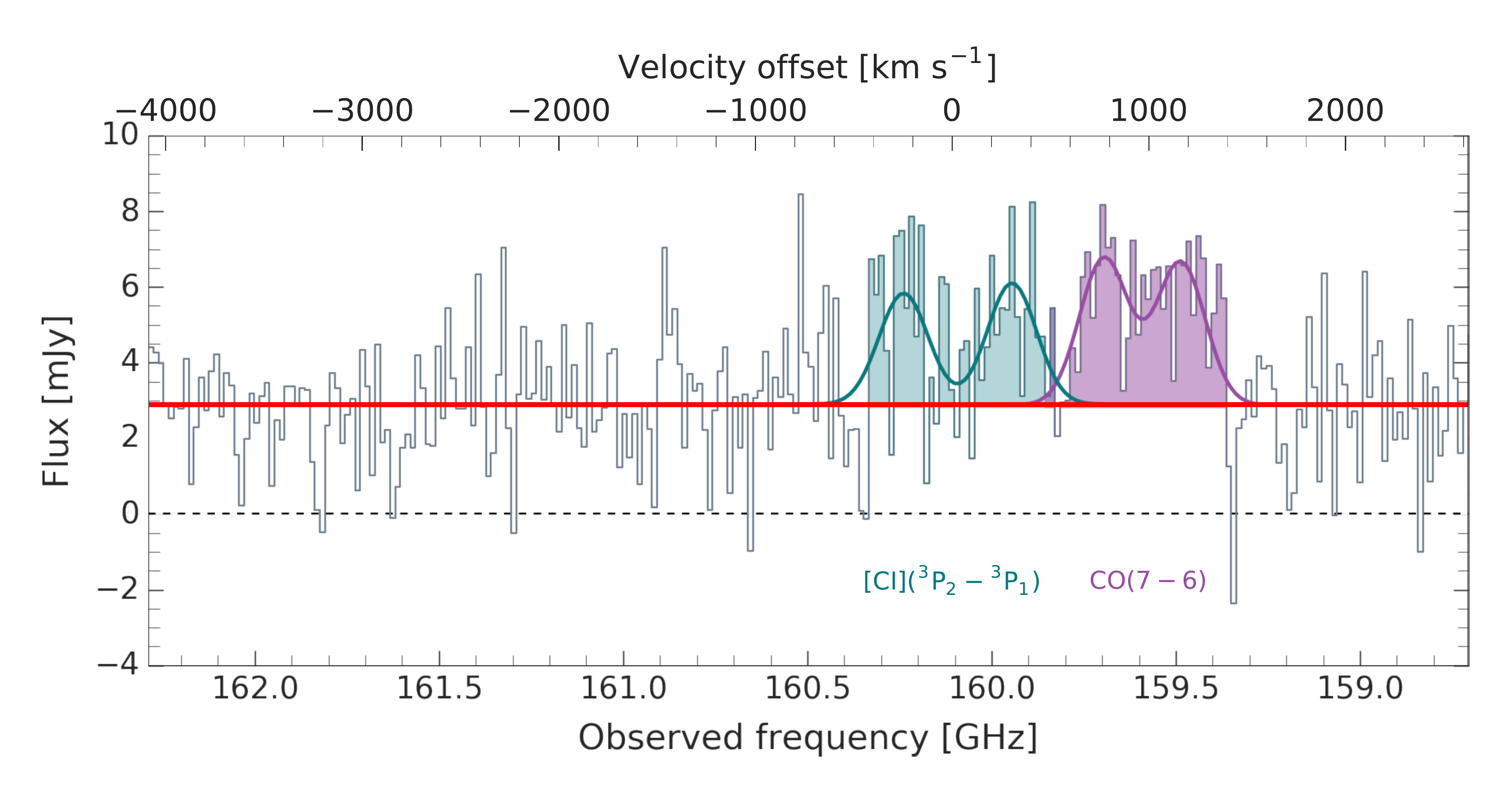}
    \caption{Extracted spectra of the \cione\ line (\textit{left}) and the 
    \citwo\ and CO(7--6) lines (\textit{right}). Both spectra are binned in steps 
    of $26\,$\si{\km\per\s}. The colored areas indicate that the velocity ranges 
    corresponding to detected line emission as labeled, which were used to 
    obtain the velocity-integrated fluxes. Blue and purple solid lines show 
    the best-fit double Gaussians, whereas the red line in each panel 
    shows the continuum level. The velocity offset in both panels is relative 
    to the expected frequency of the \ci\ lines at $z=4.0553$.} \label{fig:cico_gn20}
\end{center}
\end{figure*}

\begin{figure*}[!htb]
        \centering
        \includegraphics[width=\textwidth, trim={30 30 30 30}, clip]{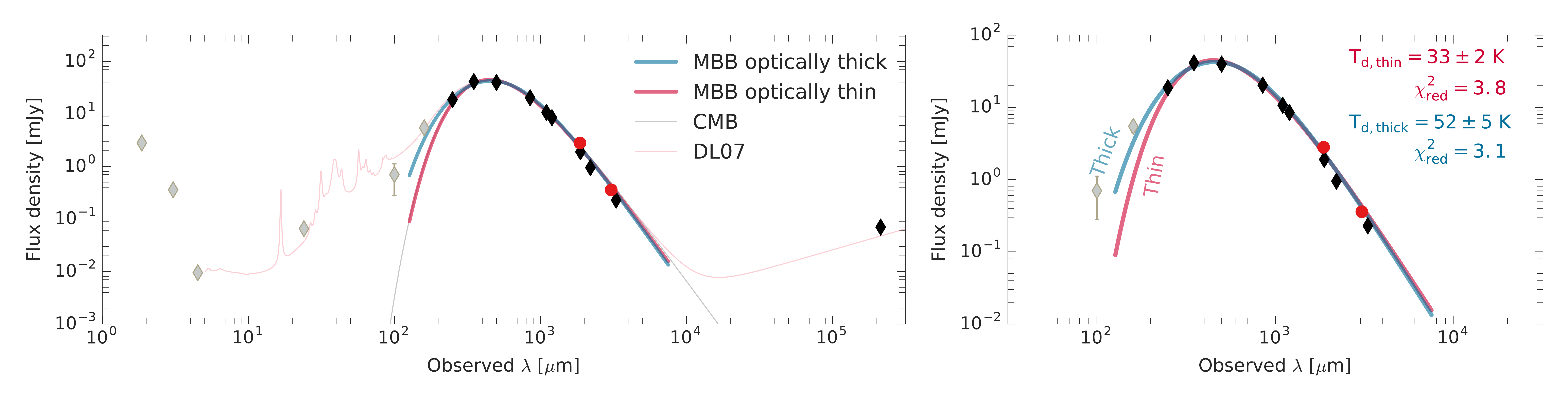}
    \caption{Mid-IR to millimeter SED of GN20 in observed 
    wavelength. \textit{Left}: we complement our new continuum 
    measurements at 1.86 and $3.05\,$mm (red points) with existing 
    photometry observations at observed and $\lambda > 160\,$\si{\micro\m} 
    and $\lambda\leq160\,$\si{\micro\m} (black and grey points, 
    respectively), where the latter is omitted from the MBB modeling.
    Blue and red lines show the best-fit single-temperature MBB 
    prescription assuming an optically thick ($\lambda_{0}=170\pm 23\,$
    \si{\micro\m}) and thin dust emission, respectively. We also 
    present the best-fit MBB model when accounting for the effect 
    of the CMB (grey curve). The solid pink line shows the best-fit 
    using the DL07 dust models, containing a diffuse ISM component and 
    dust in PDR regions. The best-fit MBB parameters are listed in 
    Table \ref{table:noema_line_obs}. \textit{Right}: A zoom-in of the 
    rest-frame FIR part of the SED of GN20 when including the 
    optically thick and thin MBB prescriptions. We note that the optically
    thick MBB model is a better macth to the photometry 
    observations at $\lambda \leq 160$ \si{\micro\m}.}
    \label{fig:GN20_SED}
\end{figure*}

The data were reduced using the GILDAS software packages CLIC and MAPPING. 
The pipeline-derived flux for our flux calibrator LKHA101 is 0.24 Jy at 97.4 GHz, 
and 0851+202 7.24 Jy at 160.1 GHz, with about 20\% absolute calibration uncertainty. 
We produced \textit{uv} tables with channel widths of 26\,\si{\km\per\s}, achieving 
an rms of 0.77 and 1.35\,\si{\mJy\per\beam} at 3\,mm and 1.86\,mm, respectively. We then 
estimated the continuum emission by averaging the line-free channels. Finally, 
we subtracted the continuum to produce the line \textit{uv} tables. The spectra were 
then extracted using the GILDAS \textit{UV\_FIT} task by assuming an intrinsic source 
size of 0.72'' (circular Gaussian FWHM), the size of the CO(4--3) line derived from the 
higher-resolution and signal-to-noise (S/N) data from \citet{Tan2014}. The beam 
sizes at 3\,mm and 1.86\,mm are $6.72"\times3.42"$ and $2.51"\times1.72"$, respectively.
The CO and \ci\ line intensity maps were produced by collapsing the \textit{uv} 
space cube according to the line widths followed by an imaging process (dirty image). 
We extracted all information directly in the \textit{uv} plane to avoid introducing 
any artifacts during the imaging process. 
We note that assuming an unresolved point-like source in the fitting leads to $\sim$20\% 
lower line fluxes and 50\% worse residuals\,\footnote{This is estimated by examining the 
total flux within one beam size aperture at the position of GN20 in the dirty image of the 
line-channel-collapsed residual data. The residual \textit{uv} data are produced by the 
GILDAS \textit{UV\_FIT} task.}.

We searched for emission lines by scanning the S/N spectra as detailed in \citet{Daddi2015}. 
The estimated continuum at $1.86\,$\si{mm} of GN20 is $2.80\pm0.13\,$\si{\mJy} and 
$0.36\pm0.04\,$\si{\mJy} at $3.05\,$\si{\mm}. The $1.86\,$\si{mm} continuum flux is 
larger than the existing measurements reported in \citet{Casey2009} ($S_{1.86 {\rm mm}}=
1.9\pm0.2\,$\si{\mJy}), for which the actual noise may have been underestimated.
On the other hand, the $3.05\,$\si{mm} continuum flux (central frequency 98.16\,GHz) is fully 
consistent with the flux reported in \citet{Tan2014} ($S_{3.3 {\rm mm}}=0.23\pm0.04\,$\si{\mJy}, 
central frequency 91.34\,GHz) when taking into account the difference in 
frequency and assuming the dust continuum decreases as $\sim \lambda^{-3.8}$.

Figure \ref{fig:cico_gn20} (\textit{left}) shows the 
\cione\ spectrum with an indication of a double-peaked structure which is more 
prominent in the \citwo\ and CO(7--6) lines (Figure \ref{fig:cico_gn20}, \textit{right}). 
We fixed the line width as derived from the brighter \citwo\ line (see Table 
\ref{table:noema_line_obs}) to estimate the \cione\ line flux. This is done for the purpose of including 
the fainter component of the \cione\ line feature which, due to the low S/N, would 
otherwise be overlooked. We detected the line with a $6.40\sigma$ significance, 
retrieving a total velocity-integrated flux of $0.70\pm0.11\,$\si{\Jy\km\per\s}.

Existing \citwo\ and CO(7--6) line observations of this target were previously 
reported as upper limits with line intensities of $<1.2\,$\si{\Jy\km\per\s} 
\citep{Casey2009}. However, our observations reveal $8.5\sigma$ and $11.0\sigma$ 
detections for the \citwo\ and CO(7--6) emission, respectively. This could 
indicate that the previous \citwo\ and CO(7--6) upper limits may  have been underestimated, 
similarly to the continuum measurement at 1.86\,mm. Figure \ref{fig:cico_gn20} (\textit{right}) 
shows the spectrum of \citwo\ and CO(7--6), where the velocity offset is relative to 
the expected frequency at $z=4.0553$. Both lines are detected and 
reveal a double-peaked structure. The total velocity-integrated flux density 
of the \citwo\ line is $1.80 \pm 0.22\,$\si{\Jy\km\per\s} with a line width 
of $949\,$\si{\km\per\s}. The observed lines indicate a redshift of $4.0536\pm0.0080  $ 
, which is consistent with previous redshift determinations from CO line measurements 
\citep{Daddi2009,Carilli2010,Carilli2011,Hodge2012, Tan2014}. The CO(7--6) flux 
measurements, along with a detailed study of the CO spectral line energy distribution (SLED), 
will be presented in a dedicated, forthcoming paper.

The total integrated flux density of each line was estimated by taking 
the product of the averaged flux density in the channels, maximizing the S/N 
and the velocity width of these channels 
\citep[see][]{Daddi2015, whitaker14}. 
We checked these non-parametric estimates against Gaussian modeling, retrieving 
fully consistent results. We proceeded with the scanning method based on 
the first approach to derive the line luminosities throughout the paper. The line 
fluxes were converted to luminosities (listed in Table \ref{table:noema_line_obs}) 
following the conversions in \citet{Solomon2005}.\vspace{-2mm}

\section{Analysis}
\subsection{The excitation temperature of neutral atomic carbon}\label{sec:ci_exc}
Our new NOEMA observations allow us to derive the excitation temperature ($T_{\rm ex}$), 
under the assumption of local thermodynamical equilibrium (LTE) and given that both 
carbon lines are optically thin. To test the validity of the latter assumption, we derived the 
optical depth of each \ci\ line following \citet{Schneider2003} (equation A.6 and A.7) by 
using the intrinsic brightness temperature of the \ci\ lines. We used the optically 
thick FIR dust results ($\tau=1$, $\lambda_{0}=170\,\mu$m, and log$(M_{\rm d}/M_{\odot})=9.31$) to 
derive the source solid angle assuming $\kappa_{850}=0.43\,$\si{\cm\squared\per\g} at 
$\lambda=850\,\mu$m yielding $\Omega_{\rm source}=2.36\times 10^{-12}\,$sr or an effective radius 
of $R_{\rm e}=1.2\,$kpc, consistent with the reported size of the rest-frame 170\,$\mu$m observations 
\citep{Hodge2015}. For the \cione\ and \citwo\ lines, we measured brightness temperatures of 
$T_{\rm b}=1.07$ and $1.02$\,K, respectively.
As the equations include the excitation temperature, we assumed for the first iteration that 
$T_{\rm ex}$ is equal to $T_{\rm d}=33-52$\,K, the derived dust temperature assuming 
optically thin and thick dust MBB prescriptions, respectively (see Section \ref{sec:sed}). 
This yields optical depths of $\tau_{\rm [CI]}=0.03-0.05$ for both \ci\ 
lines, comparable with other high-redshift galaxies \citep{Walter2011, Alaghband-Zadeh2013, Nesvadba2018}.
The excitation temperature can be derived via the formula under the assumption that 
\ci\ is thermalized, meaning that it shares the same $T_{\rm ex}$ for both levels of \ci\ \citep{Stutzki1997}:
\begin{equation}\label{eq:Tex}
    T_{\mathrm{ex}} = 38.8~\times ~{\mathrm{ln}}\left( \frac{2.11}{R} \right)^{-1}
,\end{equation}
where $R=$\lcitwo\,/\,\lci. We find $R=0.9\pm 0.2$ and $T_{\rm ex}=48.2\pm11.6\,$
\si{\kelvin}. We bootstrapped the \cione\ and \citwo\ luminosities, 
assuming normally distributed values with the observed error as the standard deviation. 
This Monte Carlo (MC) test yields a median of $T_{\rm ex}=48.2^{+15.1}_{-9.2}\,$
\si{\kelvin} (the upper and lower values are the 16th and 84th percentiles). 
Lastly, re-deriving the optical depths using the final excitation temperature 
yields $\tau_{\rm [CI]}=0.03$ for both lines, confirming that both 
\ci\ lines are optically thin\footnote{Adopting a larger size similar to 
that measured of the CO(2--1) emission \citep[$R_{\rm e}\sim4$\,kpc:][]{Carilli2010,Hodge2015} 
yields a $T_{\rm b}$ and $\tau_{\rm [CI]}$ that is $\sim 9.5$\% of values derived for the 
$R_{\rm e}=1.2\,$kpc case.}.


\subsection{Modeling of the FIR and millimeter emission}\label{sec:sed}
To further constrain the FIR and millimeter properties of GN20, we complement 
the literature observations with our new continuum flux measurements at 1.86 and 
3.05\,\si{\mm}. Existing photometry and millimeter measurements 
have already been presented in detail \citep[see][]{Magdis2012b, Tan2014} 
including photometry observations from \h\ (PACS: 100, 160\,\si{\micro\m}; 
SPIRE: 250, 350, 500\,\si{\micro\m}) and the AzTEC 1.1\,\si{\mm} 
map \citep{Perera2008}. We also include continuum measurements at 2.2, 
3.3, and 6.6\,\si{\mm} \citep{Carilli2011}, and 870\,\si{\micro\m} observations 
\citep{Hodge2015}.


\begin{table}
\centering
\caption{Derived properties of GN20.}
\begin{tabular}{lc}
\hline \hline \\ 
NOEMA observations &  \\ \hline
$I_{\rm [CI](^3P_1-^3P_0)}$ [\si{\Jy\km\per\s}] & $0.70\pm0.11^{\rm a}$ \\
$L_{\rm [CI](^3P_1-^3P_0)}^{\prime}$ [$10^{10}$ \si{\K\km\per\s\per\parsec\squared}] & $2.48\pm0.38$ \\
$I_{\rm [CI](^3P_2-^3P_1)}$ [\si{\Jy\km\per\s}] & $1.80\pm0.21$ \\
$L_{\rm [CI](^3P_2-^3P_1)}^{\prime}$ [$10^{10}$ \si{\K\km\per\s\per\parsec\squared}] & $2.33\pm0.27$ \\
$S_{\rm 3.05mm}$ [\si{\mJy}] & $0.36\pm0.04$ \\
$S_{\rm 1.86mm}$ [\si{\mJy}] & $2.80\pm 0.13$ \\ 
\hline \\
MBB best-fit solutions &  \\ \hline
$T_{\rm d, thick}$ [K] & $52\pm5$ \\
$\beta_{\rm thick}$ & $2.00\pm0.15$ \\
${\rm log}(M_{\rm d, thick}/M_{\odot})$ & $9. 31\pm0.16$ \\
${\rm log}(L_{\rm IR, thick}/L_{\odot})$ & $13.20\pm0.03$ \\
$\lambda_{0}$ [\si{\micro\m}] & $170\pm23$ \\
$T_{\rm d, thin}$ [K] & $33\pm2$ \\
$\beta_{\rm thin}$ & $1.95\pm0.11$ \\
${\rm log}(M_{\rm d, thin}/M_{\odot})$ & $9.59\pm0.10$ \\
${\rm log}(L_{\rm IR, thin}/L_{\odot})$ & $13.15\pm0.04$ \\
\hline \hline
\end{tabular}
\begin{flushleft}
$^{\rm a}$ The \cione\ line width was fixed to the best-fit of the \citwo\ line emission, 
${\rm FWHM}_{\rm [CI](2-1)}=949$\,\si{\km\per\s}.
\end{flushleft} 
\label{table:noema_line_obs}
\end{table}

We adopted three different methods to infer the FIR properties of GN20. 
First, we used the silicate-graphite-PAH models from \citet[][hereafter DL07]{Draine2007}, 
including diffuse ISM and photodissociation region (PDR) components to estimate the \lir\ 
(at 8-1000\,\si{\micro\m}), the \md, and the $\langle$U$\rangle$ by fitting the available mid-IR 
to millimeter photometry (Figure \ref{fig:GN20_SED}, \textit{left}). Since the DL07 dust models 
inherently assume that the dust emission is optically thin and do not determine a 
luminosity-weighted \td\ that is commonly used in the literature, we also considered optically 
thin and general opacity single-temperature modified blackbody (MBB) prescriptions \citep{Berta2016}.

For the general opacity MBB model, we fit the observed FIR and millimeter photometry at 
$\lambda_{\rm rest}>$ $50$\,\si{\micro\m} of GN20 (to avoid contamination from warm dust):
\begin{equation}\label{eq:MBB}
    S_{\nu} \propto (1-e^{-\tau_{\nu}}) \times B(\nu, T)
,\end{equation}
where $B(\nu_{\rm},T)$ is the Planck function, $\tau_{\nu}=(\frac{\nu}{\nu_{0}})^{\beta}$ 
is the frequency-dependent optical depth of the dust, $\nu_{0}$ is the 
frequency at which the optical depth reaches unity, and $\beta$ is the dust emissivity. 
To estimate \md, we assume a dust opacity at 850\,\si{\micro\m} of 
$\kappa_{850}=0.43\,$\si{\cm\squared \per\g} \citep{Li2001}.
In the optically thin case ($\nu_{0}\ll \nu$), the MBB prescription is reduced to:
\begin{equation}\label{eq:MBB_simp}
    S_{\nu} \propto \nu^{\beta} \times B(\nu, T)
.\end{equation}

The SED of GN20 and the best-fit prescriptions are presented in Figure \ref{fig:GN20_SED} 
and the results are listed in Table \ref{table:noema_line_obs}.
For the optically thin case, the SED fitting yields \td$=33\pm2$\,\si{\kelvin} and 
$\beta=1.9\pm 0.1$, which is consistent with the result reported in \citet{Magdis2011a} but 
considerably smaller than the $T_{\rm ex}$ derived from the \ci\ luminosity ratio 
(Section \ref{sec:ci_exc}). Accounting for the effect of the cosmic microwave background 
(CMB) on the (sub-)millimeter dust continuum emission, as detailed in \citet{daCunha2013}, 
results in consistent best-fit parameters within the uncertainties (Figure \ref{fig:GN20_SED}, left).
On the other hand, when fitting the FIR SED using a general opacity dust model 
(equation \ref{eq:MBB}), the optical depth reaches unity at a wavelength of 
$\lambda_0= c/\nu_{0} =170\pm23\,$\si{\micro\m} with a dust temperature of \td$=52\pm5\,
$\si{\kelvin}, which is fully consistent with $T_{\rm ex}$ , while recovering the same $\beta$ value 
as for the optically thin case. \vspace{-2mm}


\section{Results and discussion}
Recent works have reported a correlation between the $T_{\rm ex}$ derived from \ci\ line 
ratio and the apparent luminosity-weighted \td\ derived assuming optically thin MBB prescription 
with $\beta=2$ from resolved observations of nearby star-forming galaxies and (U)LIRGs 
\citep{Jiao2019a,Jiao2019b}. For galaxies at high-redshift, when  $T_{\rm d}$ is derived using 
the same MBB prescription, the existence of a $T_{\rm ex}-T_{\rm d}$ correlation is less clear. 
Although this is possibly due to the small sample size and lower S/N temperature estimates, which
both cause significant scatter, the high-redshift galaxies give, on average, 
$T_{\rm d} \geq T_{\rm ex}$, which is consistent with the local systems \citep{Jiao2019a,Jiao2019b,Valentino2020}.

\begin{figure}
        \centering
        \includegraphics[width=0.45\textwidth, trim={30 20 30 30}, clip]{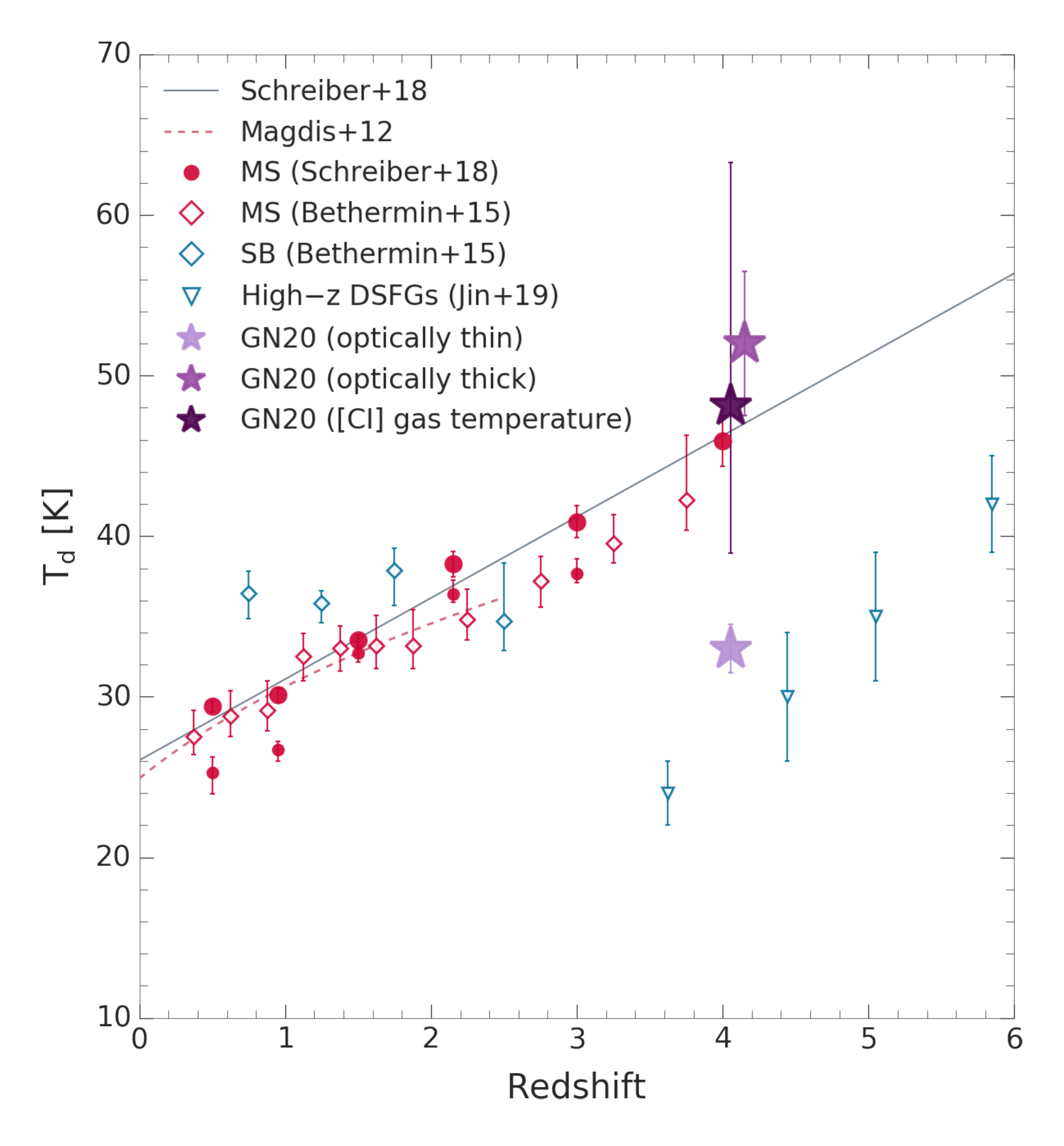}
    \caption{Evolution of \td\ as a function of redshift. 
    We include stacked MS galaxies from \protect\citet{Schreiber2018} (the small red circles 
    present the stacked galaxies in the largest mass bin with $11.0<{\rm log}(M_{\star}/M_{\odot})<11.5$ 
    whereas large red filled circles are the weighted mean of all galaxies), stacked MS and SB galaxies 
    from \protect\citet{Bethermin2015} (open red and blue symbols, respectively). For the latter, we 
    convert $\langle$U$\rangle$ to \td\ following \protect\citet{Schreiber2018}. We also include four 
    dusty SFGs from \protect\citet{Jin2019} (open blue triangles). Purple symbols depict the 
    derived $T_{\rm ex}$ of GN20 from the \ci\ luminosity ratio and from the MBB modeling assuming 
    optically thin or thick FIR dust emission.}
    \label{fig:Td_redshift}
\end{figure}

Following the same prescriptions to derive $T_{\rm ex}$ and \td\ as proposed in these studies leads to 
the observation of several curious properties for GN20. The large \ci\ line ratio yields 
$T_{\rm ex}=48.2^{+15.1}_{-9.2}\,$\si{\kelvin}, which is significantly warmer than the apparent dust 
temperature of $T_{\rm d}=33\pm 2$\,K, opposing to the general trend in the empirical 
$T_{\rm ex}-T_{\rm d}$ relation when assuming optically thin FIR dust emission. In fact, 
the \ci\ MC test predicts a 97.5\% probability of obtaining a $T_{\rm ex}$ 
above 33\,K. In Figure \ref{fig:Td_redshift}, we show the cosmic evolution of the 
luminosity-weighted dust temperature when including MS, SBs, and dusty SFGs at 
$z=0-6$ \citep{Bethermin2015, Schreiber2018, Jin2019}. The included $T_{\rm d}$ values 
from the literature are all consistent with those derived using an optically thin MBB prescription. 
We convert the mass- to luminosity-weighted $T_{\rm d}$ measurements using Eq. 6 in \citet{Schreiber2018}.
The apparent luminosity-weighted dust temperature of GN20 is similar to the 
average of main-sequence galaxies at $z\sim1.4$ \citep{Schreiber2018}, despite GN20 
being a strong starburst galaxy (${\rm SFR}=1860\pm90$\,\msol\,\si{\per\year}) and exhibiting a factor 
of $\sim6\times$ larger specific star formation rate (${\rm sSFR}=16.9$\,\si{\per\Gyr}) than $z=4$ 
MS galaxies \citep{Tan2014, Sargent2014, Jin2019}. 

Likewise, the optically thin DL07 models (assuming multi-component dust distribution)
provide similar results, yielding $\langle$U$\rangle=27.2^{+2.6}_{-2.2}$ for GN20 
\citep{Magdis2011, Magdis2012a, Tan2014}, placing it at a factor of $\sim 2.5$ times below the 
$\langle$U$\rangle-z$ relation for MS galaxies \citep{Bethermin2015, Magdis2017}. 
As a sanity check, we also converted $\langle$U$\rangle$ to \td\ following $\langle$U$\rangle= 
(T_{\rm d}/ 18.9~{\rm K})^{6.04}$ \citep{Magdis2017, Schreiber2018, Jin2019} and used the 
aforementioned conversion to obtain the luminosity-weighted dust temperature \citep{Schreiber2018}.
The inferred $T_{\rm d, DL07}=33\pm1$\,K for GN20 is fully consistent with the dust temperature 
derived from the optically thin MBB prescription. Lastly, the dust masses derived from the optically 
thin MBB and the DL07 prescriptions both lead to unphysically large $M_{\rm d}/M_{\ast}=0.04\pm0.02$ 
and $M_{\rm d}/M_{\ast}=0.05\pm0.02$, respectively, which is a factor of $\sim5\times$ higher than the predicted 
ratios for SBs based on semi-analytical models \citep{Lagos2012, Bethermin2015}.
Although the spatial offset between the optical/UV and the CO+FIR emission could indicate that 
the stellar mass is underestimated due to dust extinction, the reported dynamical mass analysis 
of GN20 \citep{Hodge2012} suggests that only a modest (if any) increase of the stellar mass can be 
allowed while still being consistent with the dynamical constraints.

Accounting for the effects of the optical depth in the SED modeling (Section \ref{sec:sed}) 
alleviates or even removes all these tensions at once. A free opacity MBB prescription 
for GN20 indicates that the FIR dust emission is optically thick up to 
$\lambda_0 = 170\pm23$\,\si{\micro\m} with an actual luminosity-weighted \td$=52\pm5$\,K 
that is similar to the $T_{\rm ex}$ from \ci\ (Figure \ref{fig:Td_redshift}), 
which is consistent with the expected dust temperature of a starburst galaxy at $z=4.05$ 
with an offset from the MS similar to GN20 \citep[Eq. 18 in][]{Schreiber2018}. The optically 
thick FIR dust temperature is also in agreement with the observed $T_{\rm ex}-T_{\rm d}$ 
relation \citep[]{Jiao2019b} of $T_{\rm ex}<T_{\rm d}$ \citep{Valentino2020}.
For a comparison with other high-$z$ starbursts, \citet{Spilker2016} report a 
$\lambda_{0}-T_{\rm d}$ correlation based on lensed starburst galaxies at $z=1.9-5.7$ 
with $\langle\lambda_{0}\rangle=140\pm40\,$\si{\micro\m}, derived using free opacity MBB 
prescription, yielding consistent results with our derived FIR properties of GN20.
Moreover, for a subsample of these galaxies, \citet{Bothwell2017} report larger 
$T_{\rm d}$ than that of the kinetic temperature ($T_{\rm kin}$) of the molecular gas 
based on \ci\ and CO molecular lines. Under the assumption of LTE, $T_{\rm kin}=T_{\rm ex}$, 
which results in $T_{\rm d}>T_{\rm ex}$, which is in agreement with previous findings.
 
As a simple check, we calculated the optical depth of the FIR dust emission similar to the 
approach described in \citet{Jin2019}, using: $\tau = \kappa \times \Sigma_{\rm dust}$ where 
$\kappa$ is the dust mass absorption coefficient from \citet{Li2001} and 
$\Sigma_{\rm dust}$ is the dust mass surface density. We derive $\Sigma_{\rm dust} 
\sim 500$\,\msol\,\si{\per\parsec\squared} assuming $R_{\rm e}\sim1.2$\,kpc 
(Section \ref{sec:ci_exc}) where $\tau\sim1$ at $\sim170$\,\si{\micro\meter},
suggesting that the dust emission is optically thick up to FIR wavelengths. 
If, indeed, the dust emission in SBs is affected by opacity effects with 
$\lambda_{0}>100\,$\si{\micro\m}, as it appears for local and high-redshift SB galaxies 
\citep[e.g.,][]{Blain2003a, Conley2011, Cox2011, Riechers2013, Simpson2017}, 
the inferred \td\ would systematically increase. This would place the SB systems 
above the \td$-z$ relation of MS galaxies at all redshifts, solving the puzzling 
observation of strong SBs being colder (or having lower $\langle$U$\rangle$) 
than MS galaxies beyond $z>2.5$ \citep{Bethermin2015}, as inferred by the optical 
thin dust models.

\begin{figure}
        \centering
        \includegraphics[width=0.45\textwidth, trim={40 30 35 35}, clip]{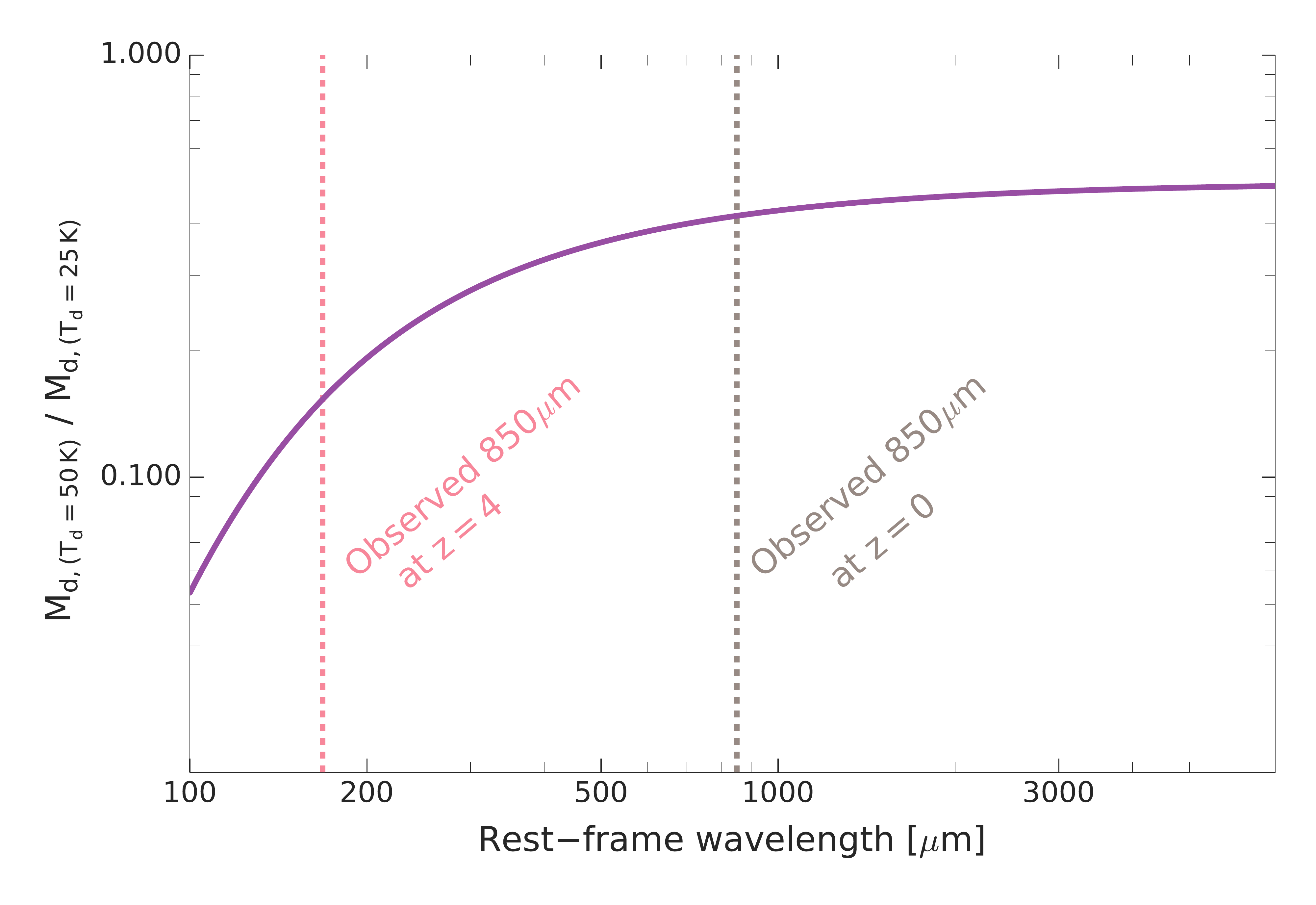}
    \caption{Comparison of the derived \md\ ratio based on the MBB 
    prescription assuming different dust temperatures (50 K compared 
    to 25 K) as a function of rest-frame wavelength. The dust continuum 
    emission at an observed 850\,\si{\micro\m} is commonly used 
    to infer \md.
    }
    \label{fig:BB_temp_ratio}
\end{figure}

An optically thick FIR dust emission will also naturally lead to lower dust 
masses. For GN20, the free opacity SED modeling results in a \md$/$\ms\ ratio of 
$0.02\pm0.01$, approaching the predicted ratios of \md$/$\ms$<0.01$ for SBs at 
$z\sim4$ \citep{Lagos2012}. The effect of the \td\ in the determination of the 
\md\ (and thus of the \mgas\ for a fixed \gdr) as a function of the rest-frame 
wavelength used to anchor the \md\ estimate is shown in Figure \ref{fig:BB_temp_ratio}. 
In the RJ tail ($\lambda_{\rm rest} \ge 500\,$\si{\micro\m}), a factor of $2\times$ 
difference in \td\ results in a factor of $\sim 2\times$ difference in \md, 
reflecting the well-known dependence of \md\ $\propto T^{-1}_{\rm d}$ in the 
optically thin limit. However, at shorter rest-frame wavelengths, the 
discrepancy between the \md\ estimates becomes considerably larger, reaching 
a factor of $\sim 5\times$ at $\lambda_{\rm rest}\sim 200\,$\si{\micro\m}.

This is a matter of caution with regard to the common approach for inferring the ISM mass 
(proportional to the \md\ and hence the \mgas) of high-$z$ galaxies from 
single-band ALMA continuum observations at observed wavelengths $850-1200\,
$\si{\micro\m} \citep[e.g.,][]{Scoville2017, Liu2019}, under the assumption of 
a fixed \gdr\ and mass-weighted \td\ of $\sim25\,$K. 
At $z>3$, such observations probe $\lambda_{\rm rest} < 300\,$
\si{\micro\m,} where moderate deviations from \td\ = 25\,K result in 
significant changes in \md\ (and thus in \mgas). Moreover, they trace a regime 
where the FIR dust emission could be optically thick. In particular, for high-$z$ 
SBs similar to GN20, an observed $850\,$\si{\micro\m} measurement probes 
$\lambda_{\rm rest} \sim 160\,$\si{\micro\m}, where the dust is likely affected by 
opacity effects. For reference, a \td\ = 25\,K versus 50\,K overestimates 
\md\ (and thus \mgas) by a factor of $\sim7\times$. We stress that the \td$=50$\,K 
measured here is luminosity-weighted and is, thus, likely to be higher than the mass-weighted \td. 

\noindent Using the $T_{\rm ex}-T_{\rm d}$ correlation to identify possible critical 
effects of the optical depth on the dust emission in extreme starbursts is potentially 
useful for settling a few issues concerning GN20. However, this relies on several assumptions 
and caveats that should be borne in mind; and alternative scenarios explaining $T_{\rm ex} 
> T_{\rm d}$ in the optically thin case might be considered. If the \ci\ line emission 
is subthermally excited, the excitation temperatures of the two \ci\ line transitions might not 
be equal as assumed under LTE. In this case, using Eq. \ref{eq:Tex} would lead to a systematically 
overestimated$T_{\rm ex}$ \citep[][but see \citealt{Israel2015} 
about the phases traced by \ci\ in extreme conditions of local starbursts]{Glover2015}. \\
Cosmic rays and turbulence could, in principle, lead to different gas and dust temperatures 
\citep{Papadopoulos2004, Bisbas2017}, assuming that the cosmic ray energy density scales with 
the SFR density  \citep{Glover2015}. An enhancement of cosmic rays is expected, thus, in 
starbursty environments, increasing the average temperature of the molecular gas, 
while at the same time, leaving the dust unaffected. An increased rate of cosmic rays in SBs would 
also lead to enhanced \ci\ emission throughout the cloud via CO destruction. However, in this case, 
models predict larger \ci\ to CO luminosity ratios in SBs than MS galaxies, which is in disagreement with 
current observations which report that the \ci/CO luminosity ratio remains roughly constant as a function of 
\lir\ and sSFR, at least on global scales \citep{Valentino2018}. A possible explanation for 
this disagreement can be caused by turbulence which can distribute \ci\ throughout the cloud, 
smoothing the \ci/CO luminosity ratio \citep{Papadopoulos2004, Bisbas2017}. As turbulence is 
expected to be dominant in regions with high cosmic ray ionization rates (i.e., in starburst 
or merger systems), it is plausible that both mechanisms are responsible for heating the 
molecular gas.

We stress that a scenario with $T_{\rm ex}>T_{\rm d}$ does not change the fact 
that the apparent dust temperature and the mean radiation field in a typical starburst galaxy at 
$z=4$ is significantly lower than that of MS galaxies at similar redshifts and that it provides an apparent 
\td\ that is in disagreement with the empirical $T_{\rm ex}-T_{\rm d}$ relation. As our study is 
based on a single galaxy, the method of using the \ci\ line ratio to distinguish between an optically 
thick or thin FIR dust solution has to be tested for the general population of high-redshift 
starbursts. However, accounting for optical depth effects at FIR wavelengths in starbursts 
similar to GN20 can mitigate several observed tensions by providing larger dust temperatures, 
in addition to lower dust masses, easing the improbable large dust to stellar mass ratios. 

\begin{acknowledgements}
We thank the anonymous referee for helpful and constructive comments which improved this paper. 
This work is based on observations carried out under project number W16DZ with the 
IRAM NOEMA Interferometer. IRAM is supported by INSU/CNRS (France), MPG (Germany) 
and IGN (Spain). IC acknowledges support from Villum Fonden research grant (13160). 
FV and GEM acknowledge the Villum Fonden research grant 13160 “Gas to stars, stars to dust: 
tracing star formation across cosmic time”, and the Carlsberg Fonden research grant CF18-0388 
“Galaxies: Rise And Death. DL acknowledges support and funding from the European Research 
Council (ERC) under the European Union's Horizon 2020 research and innovation 
programme (grant agreement No. 694343). GEM and ST acknowledge support from the ERC 
Consolidator Grant funding scheme (project ConTExt, grant number No. 648179). 
The Cosmic Dawn Center is funded by the Danish National Research Foundation.
\end{acknowledgements}\vspace{-6mm}

%
%

\bibliographystyle{aa} 
\bibliography{references}

\end{document}